# Modelling atmospheric escape and Mg$_{II}$ near-ultraviolet absorption of the highly irradiated hot Jupiter WASP-12b

*N. K. Dwivedi*[*1] , *M. L. Khodachenko*[1,2,5], *I. F. Shaikhislamov*[3], *L. Fossati*[1] , *H. Lammer*[1], *Y. Sasunov*[1,2], *A. G. Berezutskiy*[3], *I. B. Miroshnichenko*[3], *K. G. Kislyakova*[4], *C. P. Johnstone*[4], and M. Güdel*[4]

[1]*Space Research Institute, Austrian Academy of Sciences, Graz, Austria*
[2]*Skobeltsyn Institute of Nuclear Physics, Moscow State University, Moscow, Russia*
[3]*Institute of Laser Physics SB RAS, Novosibirsk, Russia*
[4]*Department of Astrophysics, University of Vienna, Vienna, Austria*
[5]*Institute of Astronomy of the Russian Academy of Sciences, 119017, Moscow Russia*

We present two-dimensional multi-fluid numerical modelling of the upper atmosphere of the hot Jupiter WASP-12b. The model includes hydrogen chemistry, and self-consistently describes the expansion of the planetary upper atmosphere and mass loss due to intensive stellar irradiation, assuming a weakly magnetized planet. We simulate the planetary upper atmosphere and its interaction with the stellar wind (SW) with and without the inclusion of tidal force and consider different XUV irradiation conditions and SW parameters. With the inclusion of tidal force, even for a fast SW, the escaping planetary material forms two streams, propagating towards and away from the star. The atmospheric escape and related mass loss rate reaching the value of $10^{12}$ gs$^{-1}$ appear to be mostly controlled by the stellar gravitational pull. We computed the column density and dynamics of Mg$_{II}$ ions considering three different sets of SW parameters and XUV fluxes. The simulations enable to compute the absorption at the position of the Mg h line and to reproduce the times of ingress and egress. In case of a slow SW and without accounting for tidal force, the high orbital velocity leads to the formation of a shock approximately in the direction of the planetary orbital motion. In this case, mass loss is proportional to the stellar XUV flux. At the same time, ignoring of tidal effects for WASP-12b is a strong simplification, so the scenario with a shock, altogether is an unrealistic one.

Keywords: planets and satellites: individual: exoplanet – planets and satellites: gaseous planets- planets and satellites: atmosphere – planets and satellites: dynamical evolution and stability – planets and satellites: atmosphere – planet-star interactions

## 1. Introduction

Hot Jupiters (HJs) are a subset of the about 4000 exoplanets known to date. They have a mass comparable to that of Jupiter and orbit their host stars at distances of less than 0.1 AU. The short orbital distance implies that these planets are highly irradiated, thus undergo extreme physical conditions, most of them connected with the interaction of the intense stellar XUV irradiation flux and stellar wind (SW) with the planetary atmosphere. Absorption of the stellar XUV flux leads to heating and expansion of the planetary upper atmosphere, which at the conditions of most HJs lead to a supersonic hydrodynamic material outflow, thus powering significant the mass loss (Lammer et al. 2003; Vidal-Madjar et al. 2003; Murray-Clay et al. 2009; Lecavelier des Etanges et al. 2010; Owen & Jackson 2012). In the most extreme cases, at very close orbital distances, the expanding atmosphere of HJs overflows the Roche lobe and the outflowing material accretes onto the star (Li et al. 2010; Fossati et al. 2010a,b; Lai et al. 2010; Matsakos et al. 2015; Pillitteri et al. 2015). The short orbital distances lead also to tidal and/or magnetic star-planet interactions that might affect the activity of the host star (Cuntz et al. 2000; Shkolnik et al. 2008; Pillitteri et al. 2015; Fossati et al. 2018).

High XUV irradiation and dense SW that impact the planetary atmosphere at close orbital distance lead also to atmospheric thermal and non-thermal escape (Khodachenko et al. 2007a, 2007b; Lammer et al. 2009). For example, Holmström et al. (2008) and Ekenbäck et al.

(2010) suggested that the escaping planetary atoms are ionized by the SW through charge-exchange with the SW protons giving origin to energetic neutral atoms (ENAs). A variety of other photochemical reactions occurs along with the radiative heating, ionization, recombination, and ENAs production (Yelle 2004; Shematovich 2012; Guo 2011, 2013). Understanding the expansion and escape occurring in the atmosphere of close-in exoplanets requires the study of the complex interconnection between SW and planetary wind (PW; Shaikhislamov et al. 2014, 2016; Khodachenko et al. 2012, 2015, 2017). However, because of the complexity of the problem, this interaction is still poorly understood.

WASP-12b is one of the most highly irradiated known transiting HJs (Hebb et al. 2009; Fossati et al. 2010b). It orbits the late F-type host star within about a day, at a distance of roughly one stellar diameter (0.0229 AU) from the stellar surface. Because of the very short orbital separation, WASP-12b is tidally locked and experiences an extreme irradiation leading to an equilibrium temperature of roughly 2500 K. Its Roche radius is only a little larger than the planetary radius (Haswell et al. 2012), thus the planet is likely to undergo the Roche lobe overflow of the upper atmospheric material followed the hydrodynamic escape driven by the intense stellar XUV irradiation.

Fossati et al. (2010a) and Haswell et al. (2012) presented the results of two near-ultraviolet (NUV) transmission spectroscopy observations conducted in 2009 and 2010 (about 6 months apart) with the Cosmic Origins Spectrographs (COS) on-board the Hubble Space Telescope (HST) aiming at the search and characterization of atmospheric escape. The observations led to the detection of transit light curves about three times deeper than in the optical, clearly indicating the presence of the escaping planetary atmospheric material outside of the Roche lobe. The observations enabled also the detection of $Fe_{II}$ and $Mg_{II}$ entailed in the outflow. The light curves obtained in the two covered spectral regions showing the largest planetary absorption indicated also the presence of a variable and early ingress, suggesting the presence of absorbing material (most likely escaped from the planet) at 4-5 planetary radii ahead of the planet.

Several hypotheses have been developed to explain the early-ingress phenomenon. One is based on the presence of mass transfer from the planet to the star. The upper atmospheric material escaped from the planet might form an accretion disk and it has been suggested the early ingress is caused by the absorption produced by an increased opacity at the position at which the stream of planetary escaped material reaches the disk. Because of conservation of angular momentum, this point would lie ahead of the planet along its orbit (Li et al. 2010; Lai et al. 2010). As the system appears to be embedded inside a torus of planetary escaped material (Haswell et al. 2012; Fossati et al. 2013; Debrecht et al. 2018), Fossati et al. (2010a) suggested that during its motion the planet might compress the gas in the torus lying ahead of it, which would then absorb the stellar light before of the planetary main transit.

However, these hypotheses do not consider the interaction of the planetary escaped material with the stellar wind, which is of great importance given that the planetary orbital velocity is supersonic (Vidotto et al. 2010). This leads to under certain condition the formation of a bow-shock ahead of the planet which opacity may be sufficient to cause the early ingress (Vidotto et al. 2010, 2011; Llama et al. 2011). Three-dimensional (3D) simulations of the planetary atmosphere and of its interaction with the stellar wind indicate the formation of a large, asymmetric gaseous envelope around the planet that is likely to reach high enough densities to produce an early ingress (Bisikalo et al. 2013, 2015). These 3D simulations are however not self-consistent and do not realistically consider the stellar irradiation, as well as some other non-adiabatic heating and cooling processes on-going in the planetary upper atmosphere, remaining therefore strongly dependent on the prescribed inner boundary conditions, close to the planet. Furthermore, they consider neutral hydrogen component only. In a series of works, we elaborated more self-consistent two-dimensional (2D) models of HJ's

PW, applied to HD209458b (Shaikhislamov et al. 2014, 2016; Khodachenko et al. 2012, 2015, 2017). Here we apply this model to the case of WASP-12b.

More specifically, we employ the 2D modeling framework of Shaikhislamov et al. (2014, 2016) and Khodachenko et al. (2012, 2015, 2017) to study atmospheric escape for a weakly magnetized WASP-12b further aiming at reproducing the observed absorption at the position of the $Mg_{II}$ h&k resonance lines and early ingress. To this end, we model the planetary atmosphere with and without considering the gravitational pull of the star, hereafter called "scenario 1" and "scenario 2", respectively. Within scenario 1, driven by tidal force the PW expands beyond the Roche lobe towards the star under the action of the stellar gravitational pull. We call this situation "captured by the star" regime. In an attempt to study the possible formation and structure of a bow-shock believed to cause the early-ingress phenomenon, we test the less realistic case ignoring tidal force (scenario 2). Within scenario 2, the background plasma velocity dominated by the Keplerian orbital speed of the planet results in a high ram pressure which stops the SW so that an ionopause and bow-shock of paraboloidal shape are formed around the planet. We call this regime of PW and SW interaction as "blown by the wind" (Shaikhislamov et al. 2016). These two scenarios aim at simulating the case of two above-mentioned specific hypotheses for the early ingress, namely the accretion stream hypothesis (scenario 1) and the bow shock hypothesis (scenario 2).

The paper is organized as follows: Section 2 describes the multi-fluid 2D hydrodynamical model. Section 3 presents the results of the simulations, which are then compared with the NUV observations in Section 4. Section 5 summaries the obtained results and presents a perspective discussion on the possible future extensions. The details regarding the applied numerical model, its equations and basic assumptions are provided in Appendix section.

## 2. Multi-fluid hydrodynamic model

To simulate the generation of the PW and its expansion, we consider a hydrogen-dominated atmosphere, irradiated by the stellar XUV, and include the effects of hydrogen ionization, recombination, different types of collisions, and basic photo-chemistry (for the list of hydrogen reactions see Table 1 in Khodachenko et al. 2015). By employing a multi-fluid approach, we overcome many of the restrictions and shortcomings of a one-fluid approach. The protons of planetary and stellar origin have different temperatures, velocities, and densities, which can be correctly considered only within a multi-fluid description. In addition, we take into account heavier species, e.g., He and Mg, to simulate the atmospheric absorption at the position of $Mg_{II}$ h resonance lines. For the minor species, we consider the following abundances: He/H=0.1 and Mg/H=$3.7 \times 10^{-5}$. Within our modeling scheme of the PW, the outflow properties are determined solely by the stellar radiation flux, the SW, and the basic system parameters, such as the orbital separation and planetary mass and radius. The major species (H and $H^+$) of planetary and stellar origin and the minor species (He and Mg) are treated as separate interacting fluids, without considering, however, of their chemistry, as it would not produce strong effect for the vast of entire hydrogen dominated material. The dynamics of Mg and its ions are described by the corresponding momentum and the ionization-recombination equations, including also the resonant double-charge-exchange with $He_{II}$. The latter may affect strongly the presence and partition of $Mg_I$ and $Mg_{II}$ around the planet up to 3-4 $R_P$ (see also Shaikhislamov et al. 2018). $Mg_{II}$ is mainly produced by the photoionization. However, $Mg_I$ and $Mg_{II}$ also react with ionized He giving $Mg_{III}$ close to the planet around 3-4 Rp. The collision coupling for $Mg_I$ with other species is calculated according to elastic cross-section $10^{-15}$ cm$^2$. $Mg_{II}$ ion is considered to be strongly coupled due to Coulomb collisions. Contribution of Mg to overall heating/cooling is neglected due to its low abundance. The basic reactions, included in the model for calculation of Mg and its ions, are listed in Appendix.

For the typical parameters of the planetary exosphere, Coulomb collisions with protons effectively couple all ions. For example, at T < $10^4$ K and $n_{H+}$ > $10^6$ cm$^{-3}$ the collisional equalization time for temperature and momentum $\tau_{Coul} \approx 2.4 T^{3/2} \left( M_i / m_p \right)^{1/2} / \left( n_{H^+} Z^3 \right)$ is about 2 s for protons, ~8 s for He$_{II}$ and ~10 s for Mg$_{II}$ ions (Braginskii 1965 Eq. (2.5.i)). This is several orders of magnitude smaller than the typical gas-dynamic timescale of the problem, which is of the order of $10^4$ s. Therefore, the minor species ions He$^+$ and Mg$_{II}$ can be supposed to be dragged by the proton flow. There is also another physical reason for the strong coupling of charged particles in the PW on the considered typical spatial scale of the problem (~$R_P$). The presence of chaotic/sporadic magnetic field in the PW affects the relative motion of the ions so that they become coupled via the magnetic field due to Lorentz force, and exchange with the momentum on a time scale of the order of Larmor period. Therefore, in spite of having negligible Coulomb collisions, the charged particles are treated as strongly coupled ones in the hot and rarefied SW as well. This suggests that there is no need of calculating the dynamics of every charged component of the plasma, and all species in the simulations are considered to have the same temperature and velocity. On the other hand, the temperature and velocity of each neutral component are calculated individually by solving corresponding energy and momentum equations. The neutral hydrogen atoms are also more or less coupled to the main flow by elastic and charge-exchange collisions. With a typical cross-section of > $10^{-16}$ cm$^{-2}$, the mean-free path at a density of $10^6$ cm$^{-3}$ is comparable to $R_P$. Besides of elastic collisions, charge-exchange ensures more efficient coupling between hydrogen atoms and protons (Shaikhislamov et al. 2016; Khodachenko et al. 2017).

As the initial condition of the simulation, we consider the planetary atmosphere composed exclusively of atomic hydrogen with a temperature of 2500 K at the inner boundary, located at the observed broadband optical planetary radius. The density of the particles at the inner boundary is fixed to provide 1 mbar thermal pressure. The choice of such a large inner boundary pressure ensures that the whole stellar XUV flux is absorbed by the atmospheric gas and the expansion of the PW does not depend on the boundary conditions (Shaikhislamov et al. 2014). The only energy input factor considered in the model is the stellar XUV flux, which we take at different values fixed at 1 AU, and then re-scaled back to the planetary orbital distance of 0.0229 AU. In recent studies the XUV fluxes chosen for the stars similar to WASP-12, e.g., WASP 13 and WASP-18 were 5.4 and 9.6 ergcm$^{-2}$s$^{-1}$, respectively at 1 AU (see Fossati et al. 2015, 2018). For simulating the escaping atmospheres of hot gas planets Salz et al. (2016) has taken the XUV flux equal to 8 ergcm$^{-2}$s$^{-1}$ for WASP-12 at 1 AU. For the same reason, in the present paper we have taken $F_{XUV}$ = 5, 10, 20 ergcm$^{-2}$s$^{-1}$. All, other parameters, such as the properties of the PW flow (e.g., temperature and velocity) and the abundances of the considered species, are self-consistently calculated in the model.

The SW is supposed not to contain neutral hydrogen, however, it is produced during the simulation because of charge exchange between the SW photons and the planetary atoms. The PW plasma satisfies the quasi-neutrality condition with equal density and thermal equilibrium of electrons and ions: $N_e = N_i$ and $T_e = T_i$. The model simulates the distributions of different species, i.e., densities of planetary protons $n_{H^+}^{pw}$, planetary atoms $n_H^{pw}$, and stellar wind protons $n_{H^+}^{sw}$ within the PW and in the area of interaction between the PW and SW. The PW-SW interaction occurs at high altitudes, while the absorption of the stellar XUV flux takes place much deeper, in the thermosphere.

The model equations are solved with an explicit second order numerical method discretized into step size of *$0.01R_p$* on a cylindrical mesh (r, z), which allows the sufficient resolution of all major physical processes, while keeping the optimal numerical part. The effect of Coriolis force is not included in our 2D model, thus restricting its range of application to sufficiently close areas around the planet, comparable to the stream thickness, i.e., $\pm 5 R_p$. The code realistically computes the fluid dynamics by setting the Courant limit to achieve stability. We take an open outer boundary condition, so that the expanding atmosphere is allowed to leave the vicinity of the planet. The details of the model can be found in Shaikhislamov et al. (2016) and Khodachenko et al. (2015, 2017) as well as in the Appendix section, which addresses the model equations, applied cross-sections, calculation of heating/ionization terms, etc. The governing physical parameters and simulation results shown below are scaled to the characteristic values of problem: temperature in units of $T_0 = 10^4$ K, speed in units of $V_0 = 9.1$ kms$^{-1}$ (ion thermal speed), distance in units of planetary radius $R_P$, and time in units of $\tau = R_P/V_0 = 3.91\,\mathrm{hr}$.

## 3. Simulation results and flow structure

In the 2D quasi-axisymmetric geometry approximation we use the gravitational potential defined by equation (A.5) in the Appendix, which has been obtained by averaging of the generalized gravitational around the planet–star line (Khodachenko et al. 2015). As briefly described in Section 1, we model two scenarios, for which in (A.5) we take into account the tidal force (realistic case) and omit it (test case). In the case of WASP-12b, this corresponds to $R_{L1} = 1.37 R_P$ and $R_{L1} = \infty$, respectively. Table 1 provides a summary of the input modeling parameters used for the simulations and gives references to the relevant figures that display the results discussed further on. Scenario 1 corresponds to the 'captured by the star' regime, in which the PW-SW interaction leads to the formation of a double stream structure with the planetary material escaping towards and away from the star (Shaikhislamov et al. 2016). The planetary escaped material could then accumulate around the star forming a torus (Haswell et al. 2012; Fossati et al. 2013), but its modeling requires a much larger simulation domain than the one considered here and a 3D geometry (Debrecht et al. 2018).

Table 1: Adopted parameters for the SW (density, temperature, and velocity) and XUV flux at 1 AU. The last column indicates the figures showing the output of corresponding simulations.

| | Density ($N_{sw}$) cm$^{-3}$ | Temperature (T) MK | Velocity ($V_{SW}$) kms$^{-1}$ | Flux ($F_{XUV}$) erg cm$^{-2}$ s$^{-1}$ at 1 A U | Figure number |
|---|---|---|---|---|---|
| Scenario 1: Captured by the star (with tidal force; fast SW) | 1x10$^4$ | 3.17 | 417 | 5 | 1,3,6(a),8(a) |
| | 1.5x10$^4$ | 3.17 | 417 | 10 | 1,3,6(c), 8(a) |
| | 1.5x10$^4$ | 3.17 | 417 | 20 | 1,3,6(e), 8(a) |
| | 3x10$^4$ | 3.17 | 417 | 5 | 2,4,6(b) |
| | 3x10$^5$ | 3.17 | 417 | 10 | 2,4,6(d) |
| | 3x10$^5$ | 3.17 | 417 | 20 | 2,4,6(f) |
| Scenario 2: Blown by the wind (No tidal force; slow SW) | 1x10$^6$ | 1.4 | 226 | 5 | 5,7(a,b),8(b) |
| | 1x10$^6$ | 1.4 | 226 | 10 | 5,7(c,d), 8(b) |
| | 1x10$^6$ | 1.4 | 226 | 20 | 5,7(e,f), 8(b) |

Scenario 2 (test case), instead, represents the 'blown by the wind' regime, in which the interaction between the SW protons and escaping PW protons leads to the formation of a parabolic ionopause and a bow-shock. The location of the bow-shock depends on the stellar XUV flux.

### 3.1 Captured by the star regime of PW-SW interaction (scenario 1)

As tidal force is taken into account, this is the more realistic case of the two, we consider. Within this scenario, we run two sets of simulations with three different XUV fluxes (5, 10, and 20 erg cm$^{-2}$s$^{-1}$) in each set, i.e., in total six simulations (see Table 1). The two sets of simulations correspond to the fast and low-density wind and the fast and dense wind, respectively. Figures 1 and 2 present the results of the simulations. The immediate result is that even a fast and dense SW (highest ram pressure) is not able to stop the PW stream driven by the tidal force in the 'captured by the star' regime. The PW stream extends beyond the Roche lobe and flows towards and away the star. Nevertheless, the thermal pressure of the surrounding SW compresses the streams on the side creating a sort of lateral ionopause boundary. The SW picks up some planetary ions as they travel close to the ionopause boundary, deflecting them towards the back of the outflow. At the same time, the majority of planetary atoms flow inside the streams and they do not reach a significant transverse velocity to cross the ionopause boundary and penetrate into the SW. We note that, because of the 2D geometry and absence of the Coriolis force, the escaping PW streams do not bend in/against the direction of the planetary orbital motion, as expected according the conservation of angular momentum (e.g., Lai et al. 2010).

Figures 3 and 4 show the PW and SW density, temperature, and velocity profiles of different constituents obtained within the simulation runs in scenario 1, with panels a, b, c, and d presenting the profiles across the star-planet line at z=2.0R$_p$, while panels e, f, g, and h show the profiles along the planet-star line (r =0).

## 3.2 Blown by the wind regime of PW-SW interaction (scenario 2)

The scenario 2 is considered only as a test case in which the tidal force is not included in the equations during the simulation runs. The main objective here was to test the formation of a bow-shock and reproduce a situation similar to that considered by Llama et al. (2010, 2011).

One of the major limitations of the quasi-axisymmetric approximation for the tidally locked system is the assumption of co-directionality for the SW plasma and XUV flux. The total SW velocity seen by the planet is the composition of the local SW velocity ($V_{SW}$) and of the Keplerian orbital velocity ($V_K$), thus $\tilde{V}_{SW} = \sqrt{V_{SW}^2 + V_K^2}$. Because of the close planetary orbital separation, the orbital component $V_K$ is very likely to be a dominant part of $\tilde{V}_{SW}$.

As WASP-12b is a tidally locked close-in exoplanet, under the limitation of the 2D geometry, in scenario 2, we considered the planet moving along the orbit with a Keplerian velocity of $V_k = 226$ kms$^{-1}$ and neglected the local SW velocity V$_{SW}$, assuming it to be smaller than V$_K$. So, the incoming SW was taken to interact with the planet along the orbital direction (see Table 1). Since the tidal force acts in the lateral direction relative to the planet orbital motion and breaks the 2D symmetry, we neglected it in scenario 2. This creates the conditions for the formation of shock in front of the planet, similar to that of Llama et al. (2011). Altogether, as compared to scenario 1, the scenario 2 considers the case of a slow SW and dense SW, taking V$_{SW}$=V$_K$ and no tidal force.

Figure 5 shows density distribution of the main interacting species, i.e., neutral hydrogen (Figs. 5a, 5d), and protons (Figs. 5b, 5e) of planetary and stellar origin in scenario 2, for two different XUV fluxes of 10 and 20 erg cm$^{-2}$s$^{-1}$ (scaled at 1 AU). Figures 5c and 5f illustrate the z-component of the velocity profiles of protons of planetary and stellar origin. Within each panel, the motion of each fluid is indicated by the streamlines. Additionally, the velocity, temperature profiles of neutral hydrogen, protons, and the density profiles of Mg$_I$ and Mg$_{II}$ are illustrated along (at r = 0: Figs. 5g, 5h, 5i) and across the planet-star line (at z = 2.0R$_p$: Figs. 5j, 5k, 5l). In scenario 2, the total SW pressure is high enough to stop the escaping PW in the direction of planetary motion and to create an ionopause and bow-shock, typical for the 'blown by the wind' regime of the PW and SW interaction (Shaikhislamov et al. 2016). In the region between the ionopause and bow-shock, the stellar protons are deflected and flow around the planet towards the tail. Figure 5 shows the presence of turbulence within the shear flow region near the ionopause boundary, with the vortices created by Kelvin-Helmholtz instabilities. A sharp change in the density distribution of both planetary atoms and ions takes place at the ionopause boundary. However, unlike the ions, planetary atoms can penetrate through the ionopause into the shocked SW region, between the ionopause and the bow-shock, which generates a difference in the density distribution of atoms and ions. Inside the ionopause, planetary atoms are strongly coupled with planetary ions via charge-exchange, which is opposite to what happens outside of the ionopause, i.e., in the SW, where the density of particles is not high enough for efficient coupling. This is indicated by the quasi-radial shape of the streamlines of the planetary atoms on the day-side, which appear to be either expanding away from the planet or trailed along the ionopause. In this region, planetary atoms interact with the stellar protons by charge-exchange.

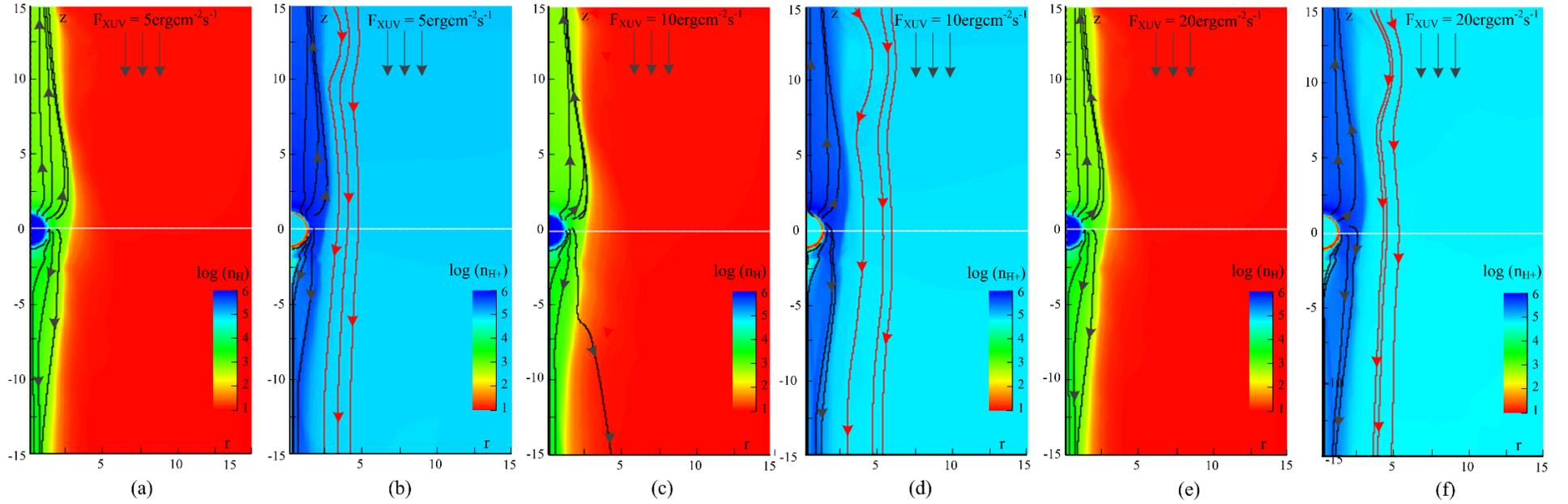

**Figure 1.** Density distribution of planetary atoms $n_H$ (a, c, e) and planetary and SW protons $n_{H+}$ (b, d, f) surrounding WASP-12b resulting from the simulations run within scenario 1. There are two panels for each considered stellar XUV flux: $F_{XUV}=5\,\mathrm{erg\,cm^{-2}\,s^{-1}}$ (a, b), $F_{XUV}=10\,\mathrm{erg\,cm^{-2}\,s^{-1}}$ (c, d), and $F_{XUV}=20\,\mathrm{erg\,cm^{-2}\,s^{-1}}$ (e, f). The planet lies at the center of the coordinate system and the star is located in the direction of the top of the panel. The motion of the corresponding components is shown by the streamlines in each panel.

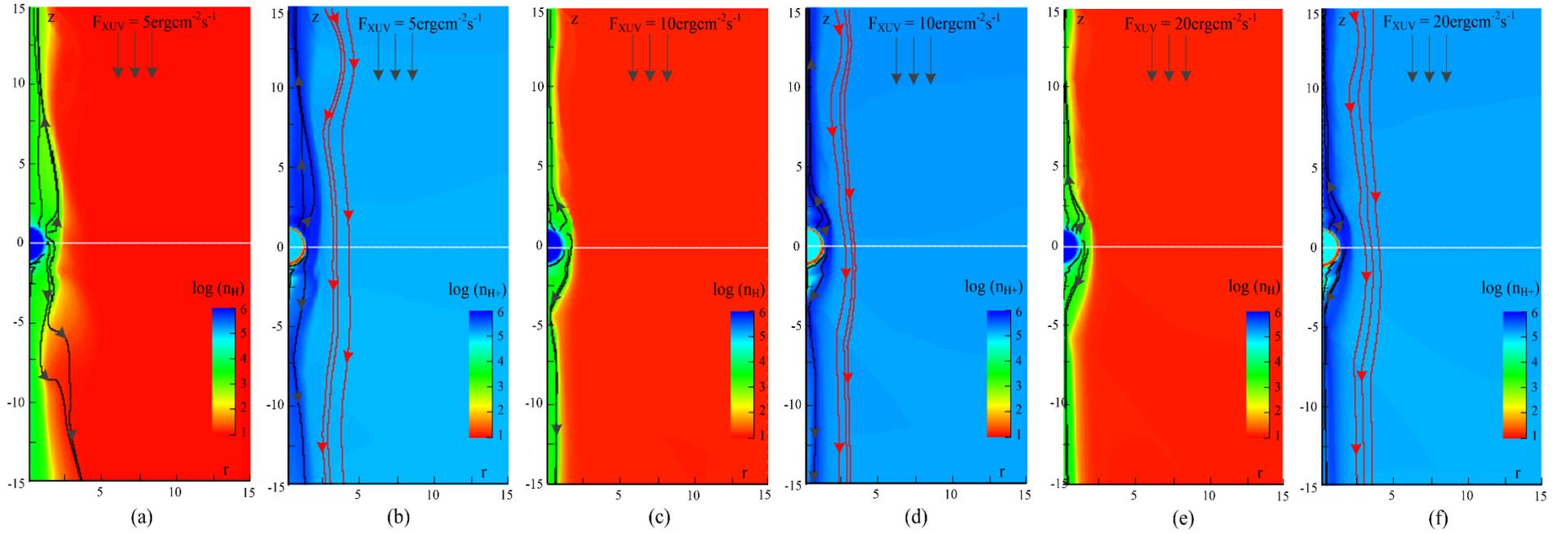

**Figure 2.** Same as Fig. 1, but for the dense SW ($N_{SW} = 3\times10^4$ cm$^{-3}$ [a, b] and $N_{SW} = 3\times10^5$ cm$^{-3}$ [c-f]) and the stellar XUV flux values of F$_{XUV}$=5ergcm$^{-2}$s$^{-1}$[a, b], F$_{XUV}$=10ergcm$^{-2}$s$^{-1}$[c, d], and F$_{XUV}$=20ergcm$^{-2}$s$^{-1}$[e, f] within the context of scenario 1.

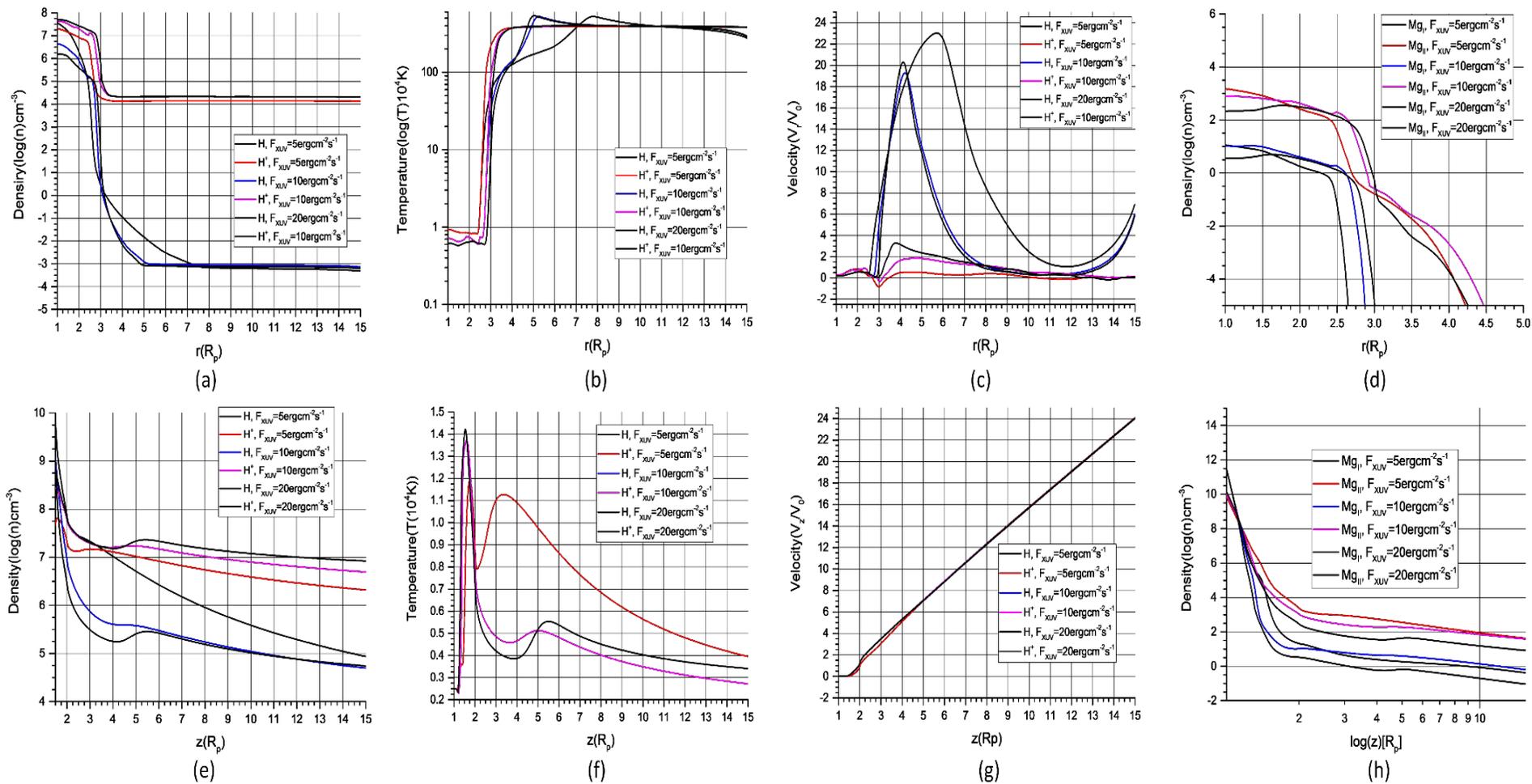

**Figure 3.** Density, temperature, and velocity profiles across (at z = 2.0$R_p$: a, b, c, d) and along (at r = 0: e, f, g) the planet-star line for the simulation runs with low SW density ($N_{SW} = 1\times10^4$ cm$^{-3}$ and $N_{SW} = 1.5\times10^4$ cm$^{-3}$ cm$^{-3}$) in scenario 1. In panel (f), the atomic hydrogen lines are overlapped with the other lines and therefore they are not visible.

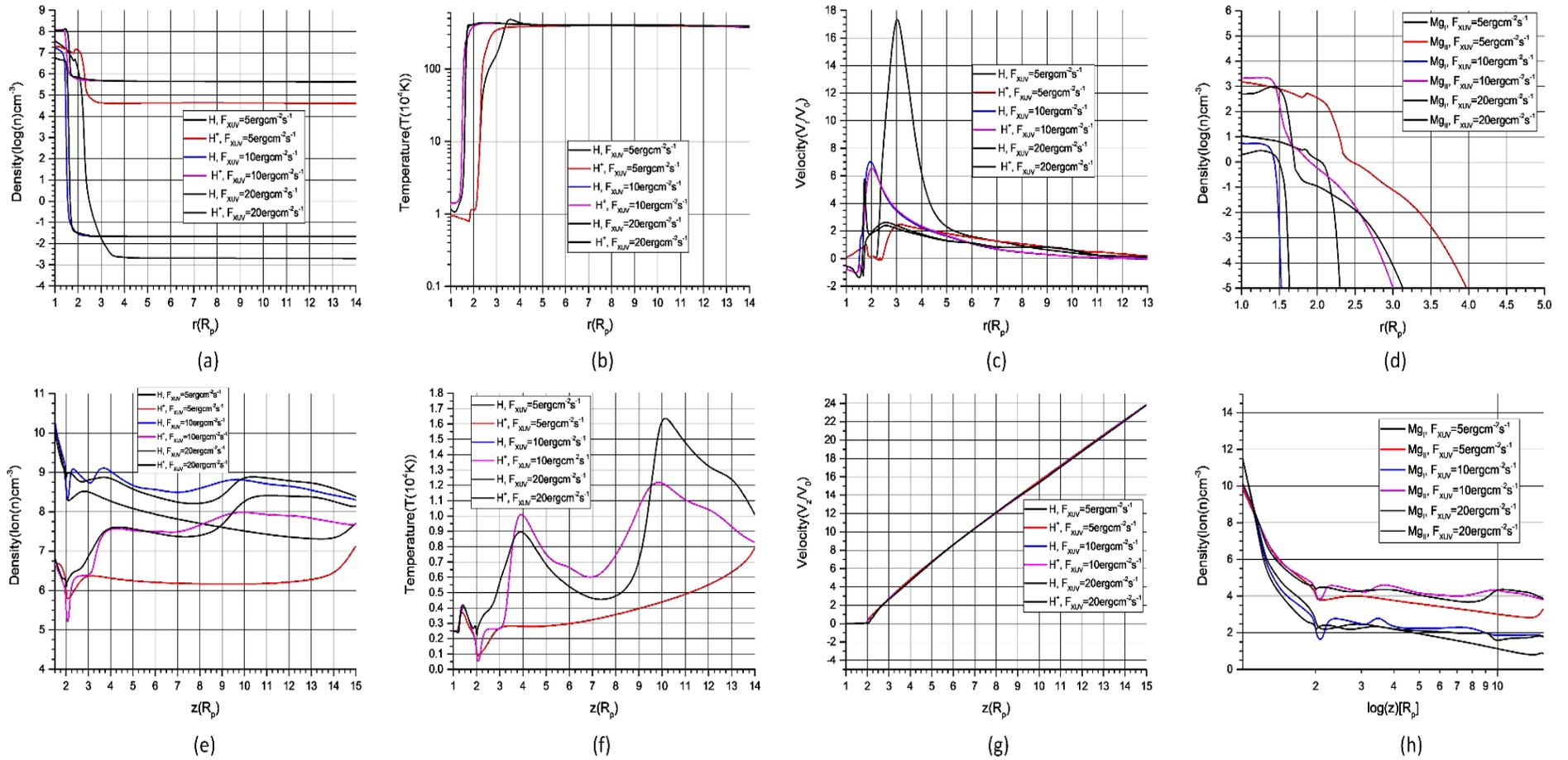

**Figure 4.** Same as Fig. 3, but for the dense SW ($N_{SW} = 3 \times 10^4$ cm$^{-3}$ and $N_{SW} = 3 \times 10^5$ cm$^{-3}$). In panel (f), the atomic hydrogen lines are overlapped with the other lines and therefore they are not visible.

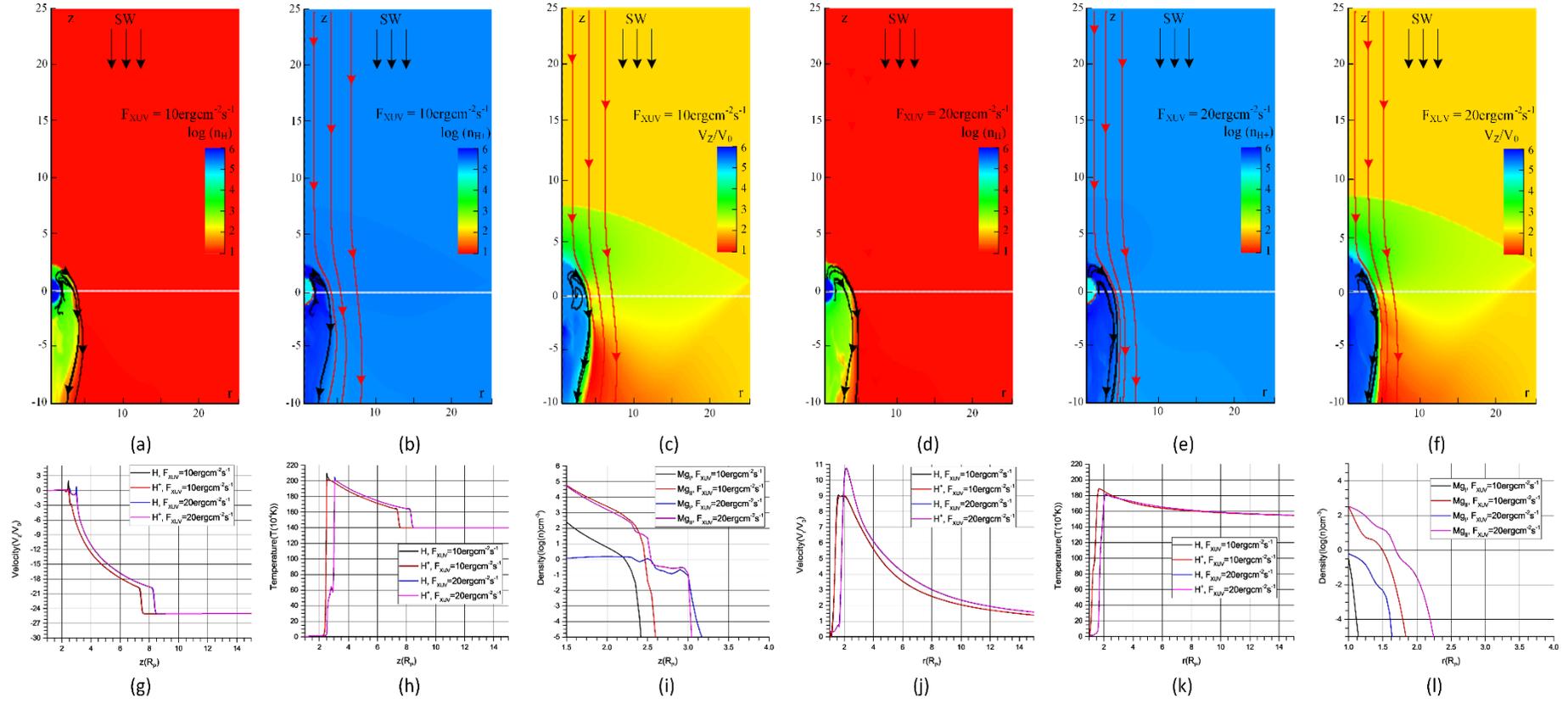

**Figure 5.** Top row: same as Fig. 1, but within the context of scenario 2 with dense SW ($N_{SW} = 1\times10^6$ cm$^{-3}$). Bottom row: velocity, temperature profiles of neutral hydrogen atoms (n$_H$), protons (n$_{H+}$), and density distributions of Mg$_I$ and Mg$_{II}$ along (at r = 0: g, h, i) and across the planet-star line (at z = 2.0R$_p$: j, k, l).

## 4. Mg$_{II}$ absorption and mass loss

Near ultraviolet observations have indicated the presence of metals in the escaping upper atmosphere of WASP-12b, particularly of Mg and Fe (Fossati et al. 2010a; Haswell et al. 2012). In this work, we also aim at roughly reproducing such observations, i.e. the transit depth, measured at the position of the Mg$_{II}$ h&k resonance lines. As Mg$_I$ atoms are rapidly photo-ionized and photo-ionization of Mg$_{II}$ is a slow process, Mg$_{II}$ should be the dominant Mg ion in the planetary atmosphere. By this, if Mg$_I$ is present in the planetary exosphere in an abundance that yields to a moderate absorption in its line, then Mg$_{II}$ should be present as well in a sufficient quantity to produce a significant (i.e., measurable) absorption. This is because of high photo-ionization energy of Mg$_{II}$, which cannot be easily quenched by recombination. As estimated by Vidal-Madjar et al. (2013), the recombination quenching of the Mg$_{II}$ stage requires an unrealistically high electron density, which itself is proportional to the ionizing XUV flux. Unlike Mg$_I$, Mg$_{II}$ ions should be present over the whole planetary exosphere producing a large resonant absorption in the $\pm 1000$ kms$^{-1}$ interval of the emission line. However, as already mentioned in Section 2, the ionization processes of Mg should also take into account the presence of He ions. In this case, Mg$_I$ undergoes a resonant charge-exchange plus ionization reaction with ionized helium, which double ionizes Mg$_I$, thus skipping the Mg$_{II}$ stage: $Mg + He^+ \rightarrow Mg^{2+} + He + e$ (DuBois 1986). At relatively high impact energies of about 40 keV, this reaction has a maximum cross-section of about $10^{-16}$ cm$^2$ (DuBois 1986). Depending on the cross-section of this reaction at very low energies, which is currently unknown, Mg$_I$ and Mg$_{II}$ might rapidly deplete when helium in the thermosphere photo-ionizes (Shaikhislamov et al., 2018). Shaikhislamov et al. (2018) investigated quantitatively these effects in the atmosphere of HD209458b and checked their impact on reproducing the observations. The account of this resonant double-charge-exchange of Mg$_I$ with He$_{II}$ affects the Mg$_I$ and Mg$_{II}$ populations in vicinity of the planet and provides their more realistic values.

One should note that no radiation pressure force acting on Mg was included in the model. It is worth mentioning here that Mg$_I$ and Mg$_{II}$ are taken as minor species and therefore their detailed chemistry is not important. The major reactions involving Mg and its ions, which are incorporated in the model, are given in the appendix. We compute spectral line absorption as in Khodachenko et al. (2017) and Shaikhislamov et al. (2018), which we describe below for completeness. We define the absorption as:

$$\text{Absorption} = 1 - \frac{I_{\text{transit},v}}{I_{\text{out},v}} = \frac{2}{R_{St}^2} \int_0^{R_{St}} \left(1 - e^{-\tau}\right) \cdot r dr \qquad (1)$$

where $\tau(V)$ is the optical depth, which is defined as:

$$\tau(V) = \int_{-L}^{L} dz \cdot n \cdot \sigma_{abs}(V, V_z, T), \tag{2}$$

and the integral is computed over the entire stellar disk.

In Eq. 2, $\sigma_{abs}$ is the absorption cross-section, which is defined as:

$$\sigma_{abs} = f_{12}\sqrt{\pi}\frac{e^2}{m_e c^2} \cdot \frac{c}{\Delta\nu_D} \cdot H = f_{12}\sqrt{\pi}\frac{e^2 \lambda_o}{m_e c^2} \cdot \sqrt{\frac{m_i c^2}{2kT}} \cdot H \tag{3}$$

where $f_{12}$ is the oscillator strength.

$$H = \frac{1}{\pi} \cdot \alpha \int \frac{e^{-y^2}}{(x-y)^2 + \alpha^2} dy \tag{4}$$

where H is the Voigt profile, which takes into account the convolution of two broadening mechanisms. Of these mechanisms, one produces a Gaussian profile (as a result of Doppler broadening) and the other one produces a Lorentzian profile (due to natural line broadening). In Eq. 4,

$$\alpha = \Delta\nu_L / 2\Delta\nu_D = \frac{\Delta\nu_L}{\nu_o}\sqrt{\frac{m_i c^2}{8kT}} \tag{5}$$

and

$$x = \frac{V - V_z}{\sqrt{2kT/m_i}} \tag{6}$$

where $\Delta\nu_L$ and $\Delta\nu_D$ are the Lorentz and Doppler line widths, respectively.

The values of the parameters required to compute the absorption, such as T, n, and $V_z$, are those computed by the simulation. The empirical value for the integration limit (L) employed to compute the optical depth along the line of sight (LOS) is taken to be *L=10R_p*. However, the computation of the absorption line of specific constituent largely depends on the presence of absorbing elements within the considered volume of integration. This means, that the area around the planet of the size of a typical ionization length, $L_{ion}$, gives the major contribution to the absorption. For example, with ionization time of ~2h and average expansion speed ~5-10 km/s, the ionization length is $L_{ion}$ ~ 1 $R_P$. The simulated $Mg_I$ and $Mg_{II}$ density distributions (Figs. 3d, h; 4d, h, 5i, l) show that these species practically vanish above ~3 $R_P$, which is also comparable to the PW stream thickness. The spiraling of the outflowing material due to the Coriolis force at such distances is still small and it should not affect strongly the calculations. This also justifies the relevance of the used 2D model approximations. Altogether, the taken sufficiently large integration length L does not really influence the absorption, while ensuring complete account of the absorbing material inside the integration domain.

Because the atmospheric temperature of WASP-12b is well above 2000 K, the Voigt integral can be fitted with an accuracy of better than 1% by the analytical expression proposed by Tasitsiomi (2013):

$$H \approx \exp(-x^2) + \frac{\alpha}{x^2\sqrt{\pi}} \cdot q(x^2) \tag{7}$$

where,

$$q = \frac{21 + x^2}{1 + x^2} \cdot z \cdot [0.1117 + z \cdot [4.421 + z \cdot (5.674 \cdot z - 9.207)]], \quad z > 0 \tag{8}$$

$$z = (x^2 - 0.855)/(x^2 + 3.42) \tag{9}$$

$$q(z<0)=0,\ q(\infty)=1,\ q_{max}=q(3.865)=1.62 \tag{10}$$

Therefore, the absorption cross-section can be represented as follows:

$$\sigma_{abs} = \sigma_{res} \cdot \sqrt{m_i \cdot 10^4 K/T} \cdot \exp(-x^2) + \sigma_{nat} \cdot \left(\frac{10\ \text{km/s}}{V-V_z}\right)^2 \cdot q(x^2) \tag{11}$$

This analytical fit explicitly shows that the absorption consists of two parts. The first part, proportional to $\sigma_{res} \approx \frac{5.9 \times 10^{-14}}{m_i}$, is due to thermal broadening and produces significant absorption at the line core (resonant process). The second part, which contains $\sigma_{nat} = 2.6 \times 10^{-18}$, is due to natural line broadening and it affects the far wings of the Lorentzian profile, thus leads to a much smaller absorption (non-resonant process). A similar formula, albeit differently normalized and without the correction factor $q(x^2)$, can be also found in Bourrier & Lecavelier des Etangs (2013).

The NUV observations reported by Fossati et al. (2010) and Haswell et al. (2012) indicate the presence of significant absorption during transit at the position of the $Mg_{II}$ h&k resonance lines, at about 2800 Å. On the basis of additional NUV observations, Nichols et al. (2015) confirmed the presence of $Mg_{II}$ absorption. Using the results of numerical simulations, in the scenarios 1 and 2 under different conditions of SW and XUV flux, the $Mg_{II}$ h line absorption profiles and transit light curves are calculated. The transit light curve is calculated by integrating the absorption over the whole line as it is detected by the observing instrument. In case of oblique line of sight (LOS), we interpolate data from grid nearest points. However in scenario 1, only the absorption line profiles of $Mg_{II}$ h are presented (Fig. 6). In scenario 2, along with the absorption line profiles of Mg$_{II}$ h, the transit light curves are also shown (Fig 7). The simulated transit light curves are further compared with the optical transit (Hebb et al. 2009). Figures 6 and 7 show the Mg$_{II}$ h line absorption profiles obtained from the simulations employing all input parameters listed in Table 1. We remind the reader that at the initial stage for all calculations we considered a Mg abundance of Mg/H=3.7x10$^{-5}$ at the inner boundary (r = R$_P$) of the simulation box and further it varied in accordance to the model calculations.

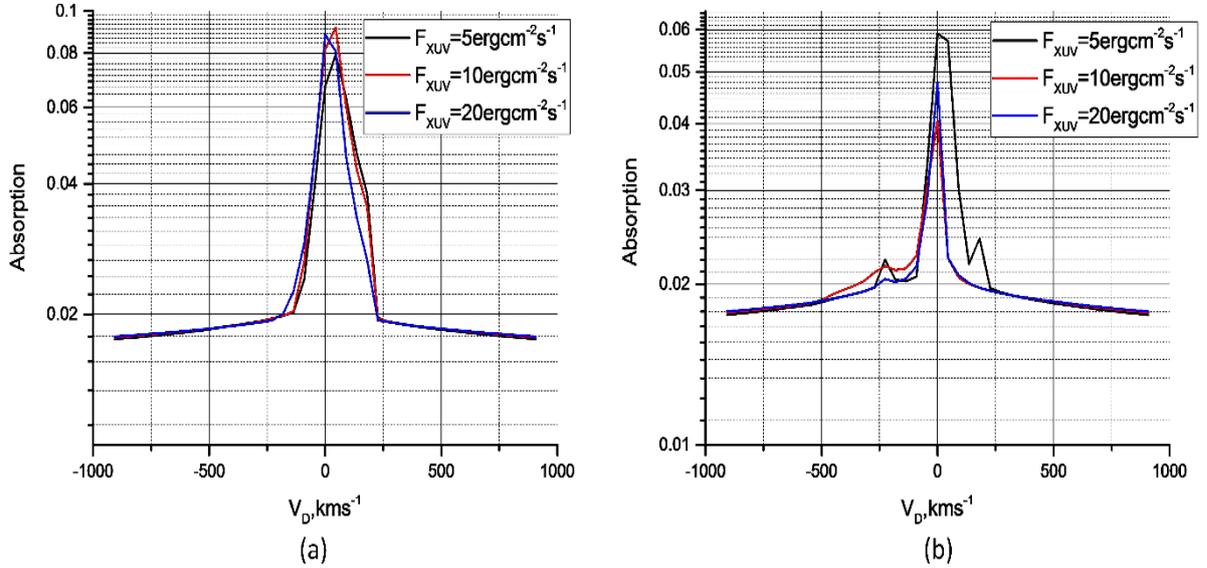

Figure 6. Absorption line profile of the $Mg_{II}$ h line in scenario 1 under three different stellar XUV flux levels: $F_{XUV}=5\,\text{ergcm}^{-2}\text{s}^{-1}$ (black), $F_{XUV}=10\,\text{ergcm}^{-2}\text{s}^{-1}$ (red), and $F_{XUV}=20\,\text{ergcm}^{-2}\text{s}^{-1}$ (blue), Panel (a): rarified SW ($N_{SW}=1\times10^4\ \text{cm}^{-3}$ and $N_{SW}=1.5\times10^4\ \text{cm}^{-3}$); Panel (b): dense SW ($N_{SW}=3\times10^4\ \text{cm}^{-3}$ and $N_{SW}=3\times10^5\ \text{cm}^{-3}$).

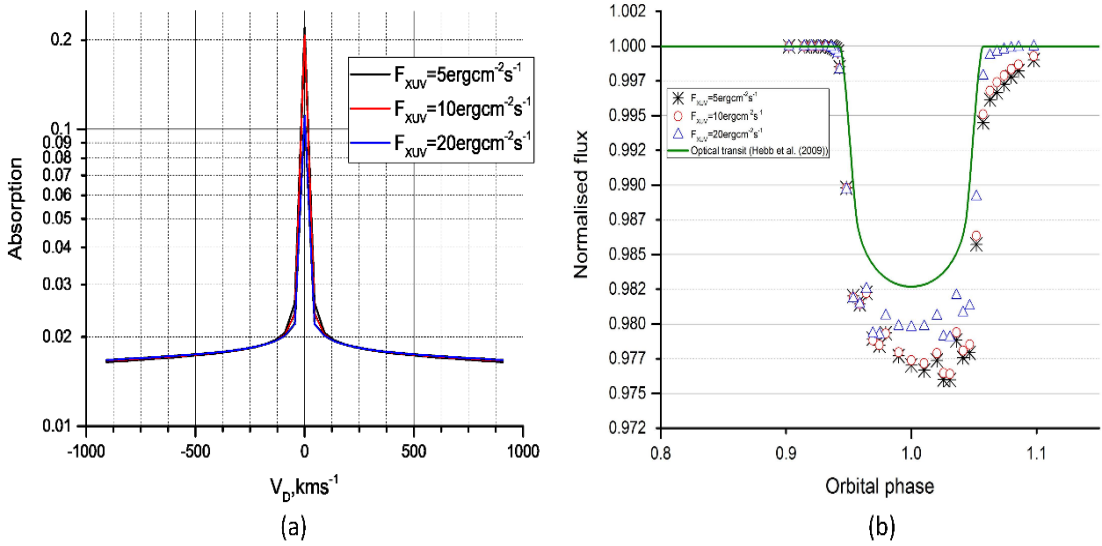

**Figure 7.** Absorption line profile of the Mg$_{II}$ h line (a) and transit light curves (b) derived from the simulations for the scenario 2 under over-dense SW ($N_{SW} = 1 \times 10^6$ cm$^{-3}$) and three different stellar XUV flux levels: F$_{XUV}$=5ergcm$^{-2}$s$^{-1}$ (black), F$_{XUV}$=10ergcm$^{-2}$s$^{-1}$ (red), and F$_{XUV}$=20ergcm$^{-2}$s$^{-1}$ (blue). The green line in panel (b) is the optical transit (Hebb et al. 2009).

In scenario 1, we find that the SW plays a much greater role in shaping the absorption line profile in comparison to the stellar XUV flux. Figure 6 shows that a denser SW leads to a smaller absorption that is caused by a more compact stream of escaping PW. Within scenario 1, we obtain the closest match to the observed Mg$_{II}$ absorption (~about 4% absorption). This is achieved with $N_{SW} = 3 \times 10^4, 3 \times 10^5$ cm$^{-3}$, $V_{SW} = 417$ kms$^{-1}$, $T_{SW} = 3.17$ MK. As it can be clearly seen in Figure 6, under the rarified SW condition the total absorption in the Mg$_{II}$ h lines (panel a) has higher values than those computed for the dense SW (panel b). Our simulations do not lead to the formation of an early ingress in scenario 1, regardless of the adopted SW and XUV flux values, as the structure of the escaping material stream is rather symmetric. This is obviously connected to the nature of the 2D model. Contrary within scenario 2 ('blown by the wind' regime), considering a slow SW and ignoring tidal force, we observe the formation of a bow-shock ahead of the planet and obtain strong absorption, independently of the stellar input parameters. The corresponding computed line profiles and transit light curves are shown in Figure 7 for three different XUV fluxes and SW parameters according to Table 1. We observe that the ingress and egress start approximately at the same time as those in optical photometry. The only differences consist in the depth of the transits.

Atmospheric escape plays a crucial role in the evolution of planets. Therefore, we estimate the planetary mass loss rate and show its variation in dependence of the model parameters and scenarios for the simulated runs in Fig. 8. Within scenario 1, in which the mass loss rate is significantly affected by the tidal force, we found the mass loss rate increased by 10-15 times compared to that computed within scenario 2, in which the mass loss rate is influenced only by the XUV radiation flux. Note that the increasing stage seen in all lines in Fig. 8 corresponds to the numerical model takeoff phase after a simulation start and before achieving the quasi-steady-state with the meaningful values. The average saturated values for the mass loss rate are listed in Table 2.

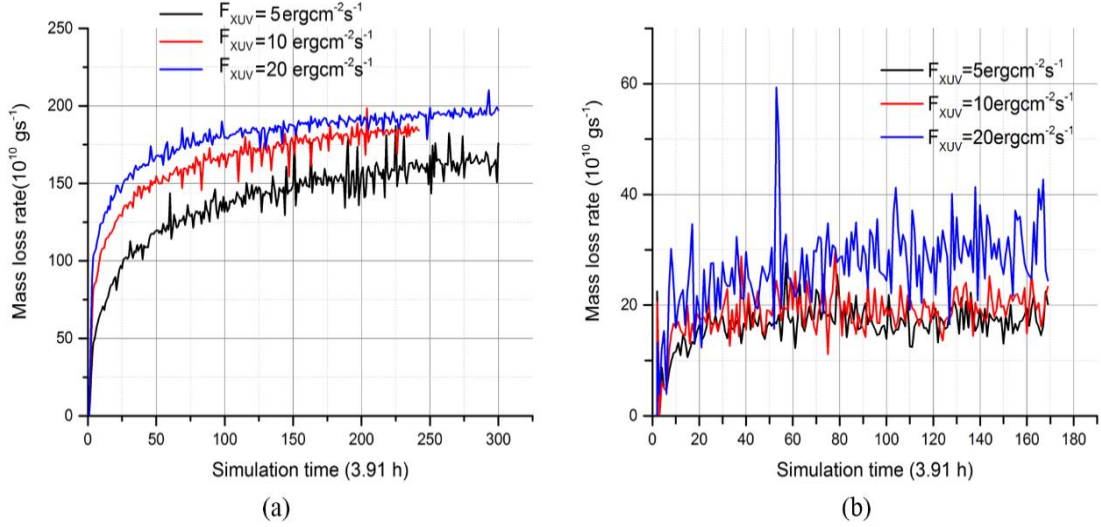

**Figure 8.** The simulated mass loss rate as a function of time (in units of $\tau = 3.91\,\mathrm{hr}$) obtained for the scenario 1 (left) and 2 (right).

**Table 2:** XUV flux, SW parameters and Mass loss rates (saturated average value) for scenarios 1 and 2

| | $F_{XUV}$ (ergcm$^{-2}$s$^{-1}$) | $N_{SW}$ (cm$^{-3}$) | $V_{SW}$ (kms$^{-1}$) | $T_{SW}$ (MK) | Mass loss rate ($10^{10}$ gs$^{-1}$) |
|---|---|---|---|---|---|
| **Scenario 1: Captured by the star (with tidal force; fast SW)** | 5 | $1\times10^4$ | 417 | 3.17 | 145.38 |
| | 10 | $1.5\times10^4$ | 417 | 3.17 | 169.95 |
| | 20 | $1.5\times10^4$ | 417 | 3.17 | 186.53 |
| | 5 | $3\times10^4$ | 417 | 3.17 | 197.64 |
| | 10 | $3\times10^5$ | 417 | 3.17 | 222.98 |
| | 20 | $3\times10^5$ | 417 | 3.17 | 225.29 |
| **Scenario 2: Blown by the wind (no tidal force; slow SW)** | 5 | $1\times10^6$ | 226 | 1.4 | 9.91 |
| | 10 | $1\times10^6$ | 226 | 1.4 | 11.91 |
| | 20 | $1\times10^6$ | 226 | 1.4 | 20.49 |

## 5. Conclusions

We presented the results of a number of simulation runs performed with different XUV irradiation fluxes and SW conditions for two scenarios. In the most realistic scenario 1, i.e., 'captured by the star' regime of the PW and SW interaction, taking into account all major physical effects, the SW total pressure cannot balance the stellar gravitational pull on the escaping material of the planetary atmosphere. This case is analogous to the one discussed in Tremblin & Chiang (2013). The effect of large-scale spiraling of the escaping PW stream due to Coriolis force is however not covered by the model, while the plasma and especially the absorbing Mg$_{II}$ conditions within the distances range ~3 $R_P$ comparable to the stream width are still reproduced correctly. In a more simplified case, within scenario 2, with 'blown by the wind' regime, neglecting the tidal force, pressure-dominated plasmasphere is created which is bounded by the ionopause with a stand-off pressure balance point located between $2R_P - 3R_P$. The supersonic velocity of orbital motion in scenario 2 results in the formation of a bow-shock ahead of the planet at $5R_P - 8R_P$ depending on the F$_{XUV}$ irradiation level, while the PW flow is redirected towards the tail. We also found that the strongly coupled neutral and ionized components of the expanding PW decouple at the ionopause boundary, where the planetary and stellar proton flows are being stopped by each other. However, the neutrals are accelerated by the pressure gradient and penetrate into the SW. The multi-fluid simulations performed in the present paper endorse the idea proposed in our previous works (Shaikhislamov et al. 2014, 2016; Khodachenko et al. 2015, 2017) to estimate the transit absorption spectra on the basis of numerically modelled nearby plasma environments of a planet. The bulk motion of the escaping planetary material with a velocity of several tens of kms$^{-1}$, drags heavy elements, that sufficiently increase the contribution of the resonant thermal line broadening mechanism, which dominates in the absorption of corresponding stellar lines.

The model in particular enables the simulation of Mg$_{II}$ resonant and absorption line profiles within the range of Doppler shift up to $\pm 1000$ kms$^{-1}$, measured also during the transits of WASP-12b. In addition, we have estimated the upper atmospheric mass loss rates in both scenarios with the corresponding regimes of PW and SW interaction. The Mg$_{II}$ absorption lines are found to be dependent on XUV flux in scenario 2 (blown by the wind), whereas in scenario 1 (captured by the star), the influence of tidal force and SW conditions dominate. In scenario 1, the simulated Mg$_{II}$ absorption lines show a variation between 4-10% with the higher values of maximum achieved under the rarified SW condition. For dense SW, the simulated absorption lines have the maximum at 4-6%. In Scenario 2, a substantial enhancement (up to 22% - 25%) in the maximum value of Mg$_{II}$ absorption is observed depending on the value of F$_{XUV}$, changed from 5ergcm$^{-2}$s$^{-1}$ to 10ergcm$^{-2}$s$^{-1}$. However, the maximum value of Mg$_{II}$ absorption decreases down to 12% for F$_{XUV}$ = 20ergcm$^{-2}$s$^{-1}$, due to the higher ionization degree of Mg for intense XUV irradiation levels. We have also computed the transit light curves in the case of the 'blown by the wind' regime of PW and SW interaction. We observed that the simulated light curves with F$_{XUV}$ = 5, 10, and 20 ergcm$^{-2}$s$^{-1}$ start exactly at the optical transit in scenario 2. This is because of the quickly decreasing Mg$_{II}$ density, still remains symmetrically distributed in the vicinity of the planet caused by high ionization degree of Mg at the considered F$_{XUV}$ values. Previous studies often appealed to the formation of a bow-shock under the prescribed PW conditions; however, in the simulation runs, performed with a self-consistent PW, in accordance with the realistic XUV fluxes and different SW parameters, no bow-shock was formed in the case of 'captured by the star' regime of PW and SW interaction. A complete understanding of the details of shock formation and a possibility of material density enhancement in front of WASP-12b, and its implication on the early ingress is complex problem, which requires self-consistent numerical simulations in a three-dimensional geometry. The meticulous self-consistent multi-fluid simulation of expanding atmosphere of WASP-12b and injection of planetary atoms into the SW with a subsequent absorption of Mg$_{II}$ is the major novelty provided by our model and the present investigation. Furthermore, one of the most interesting results of the present study lies in the 'captured by the star' regime, i.e., scenario 1, which might take place for the real WASP-12b, when the tidally driven escaping planetary atmosphere material achieves high velocities and forms a double stream structure.

**Acknowledgement**

*This work is supported by the Austrian Science Foundation (FWF) project I2939-N27, Austrian Agency for International Cooperation in Education and Research project No. IN 05/2018, and the Russian Science Foundation grant № 18-12-00080. MLK, HL, MG, KGK, and CPJ acknowledge FWF-NFN projects S11606-N16, S11604-N16, and S11607-N16. Parallel computing simulations, the key for this study, have been performed at Computation Center of Novosibirsk State University, SB RAS Siberian Supercomputer Center, and Supercomputing Center of the Lomonosov Moscow State University.*

# **Appendix A**

**The Model Equations**

To simulate the dynamics of the escaping hydrogen dominated atmosphere of WASP-12b and its interaction with the SW, we employ a 2D axially symmetric hydrodynamic multi-fluid numerical model in the cylindrical coordinate system with the symmetry axis, *z*, taken along the planet-star line and the reference frame attached to the planet. Such geometry is well suited for the simulation of tidally locked non-magnetized planets, with the stellar illumination coming from only one (dayside) direction. Here we provide some details regarding the numerical model and its basic equations. The model code solves numerically the continuity, momentum and energy equations for separate components, which can be written in the following form (Shaikhislamov et al. 2016):

$$\frac{\partial}{\partial t}n_j + \nabla(\mathbf{V}_j n_j) = N_{XUV,j} + N_{exh,j} \quad (A.1)$$

$$m_j \frac{\partial}{\partial t}\mathbf{V}_j + m(\mathbf{V}_j \nabla)\mathbf{V}_j = -\frac{1}{n_j}\nabla n_j kT_j - \frac{z_j}{n_e}\nabla n_e kT_e - m_j \nabla U - m_j \sum_i C_{ji}^\nu (\mathbf{V}_j - \mathbf{V}_i) \quad (A.2)$$

$$\frac{\partial}{\partial t}T_j + (\mathbf{V}_j \nabla)T_j + (\gamma - 1)T_j \nabla \mathbf{V}_j = W_{XUV,j} - \sum_i C_{ji}^T (T_j - T_i) \quad (A.3)$$

The main processes, responsible for the transformation between neutral and ionized particles are photo-ionization, electron impact ionization, and dielectronic recombination. These are included in the term $N_{XUV,j}$ in continuity equation (A.1) and are applied for all species. Photo-ionization also results in a strong heating of planetary material by the produced photo-electrons. The corresponding term $W_{XUV,j}$ in the energy equation (A.3) (addressed in Shaikhislamov et al. 2014, Khodachenko et al. 2015) is derived by integration of the stellar XUV spectrum. For the solar type host star such as WASP-12b we use here as a proxy the spectrum of the Sun, compiled by (Tobiska 1993) and covering the range 10–912 Å, binned by 1 Å. The spectrum is based on measurements of solar radiation under moderate activity conditions with proxy index $P_{10.7} = 148$. It is assumed that the energy released in the form of photo-electrons is rapidly and equally re-distributed between all locally present particles with efficiency of $\eta_h = 0.5$. This is a commonly used assumption, which we adopted on the basis of qualitative analysis (Shaikhislamov et al. 2014). The heating term includes also energy loss due to excitation and ionization of hydrogen atom and in simplified form can be written as:

$$W_{XUV,j} = (\gamma - 1) \cdot n_a \left[ \langle (\hbar \nu - E_{ion}) \sigma_{XUV} F_{XUV} \rangle - n_e \upsilon_{Te} (E_{21} \sigma_{12} + E_{ion} \sigma_{ion}) \right] \quad (A.4)$$

Another kind of important interaction between the considered particle populations is resonant charge-exchange collisions. Indeed, charge-exchange has the cross-section of about $\sigma_{exc} = 6 \cdot 10^{-15}$ cm$^2$ at low energies, which is an order of magnitude larger than the elastic collision cross-section. Experimental data on the differential cross-sections can be found, for example, in Lindsay & Stebbings (2005). When planetary atoms and protons slip relative each other, because they have different thermal pressure profiles and protons feel electron pressure while atoms do not, the charge-exchange between them leads to velocity and temperature interchange. We describe this process with a collision rate $C_{ji}^\upsilon$ where upper index indicate the value being interchanged. For example, in momentum equation for planetary protons it looks as $C_{H^+H}^\upsilon = n_H^{pw} \sigma_{exh} \upsilon$, where the interaction velocity $\upsilon \approx \sqrt{V_{Ti}^2 + V_{Tj}^2 + (V_j - V_i)^2}$ depends in general on thermal and relative velocities of the interacting fluids (in the considered example, protons and neutral atoms). More precise expressions for charge-exchange terms in the continuity, momentum and energy equations, obtained by averaging of the collision operator over the Maxwell distribution (e.g., Meier & Shumlak 2012), differ from those used in our work by an order of unity coefficients or by small additional terms, which are inessential for the present study.

In above equations, the standard cross section of the governing processes were considered, e.g., $\sigma_{XUV} = 6.3 \times 10^{-18} \cdot (\lambda/\lambda_{thr})^3$ cm$^2$, wavelength dependent ionization cross section by XUV; $\sigma_{ion} = 4 \times 10^{-16} \cdot e^{-E_{ion}/T} \cdot T^{-1}$ cm$^2$, ionization cross section due to electron impact, $\sigma_{rec} = 6.7 \times 10^{-21} T^{-3/2}$ cm$^2$, cross section of the recombination with the electron; $\sigma_{12} = \sigma_{21} \cdot e^{-E_{21}/T}$ cm$^2$ and $\sigma_{21} = 7 \times 10^{-16} \cdot T^{-1}$ cm$^2$ as hydrogen excitation and de-excitation respectively; and Lyα absorption as $\sigma_{L\alpha} \sim (10^{-15}...10^{-14}) T^{-1/2}$ cm$^2$ with the temperature scaled in units of characteristic temperature, i.e., $\tilde{T}_c = 10^4$ K, except the exponent power indices in the expressions for $\sigma_{ion}$ and $\sigma_{12}$, where temperature was considered in the energy units.

For the typical parameters of the planetary plasmasphere, Coulomb collisions with protons effectively couple the minor species ions. For example, at T<$10^4$ K and $n_{H^+} > 10^6$ cm$^{-3}$ the collisional equalization time for temperature and momentum $(C_{H+,j}^\upsilon)^{-1} = \tau_{Coul} \approx 2.4 T^{3/2} (M_i/m_p)^{1/2} / (n_{H^+} Z^3)$ is about 2 s for protons and aprroximately 8 s and 10 s for He$_{II}$ and Mg$_{II}$ ions, respectively (Braginskii, 1965). This is several orders of magnitude less than the typical gas-dynamic time scale of the problem, treated here, which is of the order of $10^4$ s. There is also another physical reason for the strong coupling of charged particles in the PW on the considered typical spatial scale of the problem, that is about R$_p$ ($\sim 10^{10}$ cm). The chaotic/sporadic magnetic field, which is present in the PW, and affects the relative motion of the ions so that they become coupled via the magnetic field due to the Lorentz force, and exchange with the momentum on a time scale of the Larmor period. For the same reason, the charged particles can be treated as strongly coupled ones in the hot and rarefied SW as well, even in spite of the fact that Coulomb collisions are negligible there. Therefore, there is no need to calculate the dynamics of every charged component of the plasma fluid species, and we assume in the simulations all of them to have the same temperature and velocity. On the other hand, the temperature and velocity of each neutral component are calculated individually by solving the corresponding energy and momentum equations. The neutral hydrogen atoms are more or less coupled to the main flow also by elastic collisions. With a typical cross-section of >$10^{-16}$ cm$^2$, the mean-free path at a density of $10^6$ cm$^{-3}$ is comparable to $R_p$. Besides elastic collisions, charge-exchange ensures more efficient coupling between hydrogen atoms and protons (Shaikhislamov et al. 2016; Khodachenko et al. 2017).

In view of a strong coupling of the proton fluids of the planetary and stellar origin, they have the same velocity and temperature whenever they mix. Therefore, the numerical model contains a physically equivalent single proton fluid which is a subject to two different boundary conditions – at the planet's surface (zero density and velocity) and at the outer boundary (SW parameters). Account of molecular hydrogen and the corresponding molecular ions $H_2^+, H_3^+$, like in Shaikhislamov et al. (2018), allows more accurate treatment of the inner regions of the planetary thermosphere, but as it was shown in Shaikhislamov et al. (2014) and Khodachenko et al. (2015), as well as in Yelle (2004) and Koskinen et al. (2007), the components $H_2, H_2^+, H_3^+$ can exist only at very low heights $< 0.1 R_P$ in a relatively dense thermosphere of a hot-Jupiter-type planet. Therefore, they do not influence significantly the parameters of the PW flow in the region of its interaction with the SW, and consequently, does not affect the dynamics of the absorbing agent (i.e., Mg) in the vicinity of the planet. In fact, the interaction between the PW and SW, which takes place at higher altitudes of several planetary radii $R_P$, involves only the dominating protons and hydrogen atoms. For these reasons in the present paper we disregard molecular species, and assume that the initial atmosphere of the WASP-12b consists of only hydrogen atoms and Helium in a barometric equilibrium.

The model admits also inclusion of minor species, which are described as separate fluids by the corresponding momentum and continuity equations. The population of different ionization states for each element is calculated assuming the specific photo-ionization (Verner et al. 1996) and recombination rates (Le Teuff et al. 2000, Nahar & Pradhan 1997). Note, that we do not consider chemical reactions among the different minor species, whereas the list of modeled hydrogen reactions is presented in Khodachenko et al. (2015), and it is practically the same as that used in other aeronomy models (e.g., García Muñoz 2007; Koskinen et al. 2007), except that no reactions between He and H are considered. In the present modeling of WASP-12b besides of the dominating hydrogen, we consider the helium (He) and magnesium (Mg) components of the planetary origin taken with the typical solar system abundances (relative to hydrogen) He/H=0.1, and Mg/H=3.7·$10^{-5}$, respectively (Anders & Grevesse 1989). At typical temperatures of less than $10^4$ K and electron densities of less than $10^9$ cm$^{-3}$, Mg$_{II}$ prevails over Mg$_I$ (Vidal-Madjar et al. 2013). We also include the resonant double charge-exchange with ionized helium which double ionizes Mg$_I$, while skipping the Mg$_{II}$ stage: Mg + He$^+$ = Mg$^{2+}$ + He + e (DuBois 1986). The relevance of this reaction is studied in (Shaikhislamov et al. 2018). Mg$_I$ and Mg$_{II}$ are taken as minor species and therefore the detailed chemistry of Mg$_I$ and Mg$_{II}$ is not important, however, following reactions are considered in the paper:

a) Photoionization: $Mg_I + h\upsilon \rightarrow Mg_{II} + e$

b) Recombination: $Mg_{II} + e \rightarrow Mg_I$

c) Charge-exchange: $Mg_I + H^+ \rightarrow Mg_{II} + H$, $\quad Mg_{II} + H^+ \rightarrow Mg_{III} + H$, and $Mg_{II} + He^+ \rightarrow Mg_{III} + He$

d) Double charge-exchange: $Mg_I + He^+ \rightarrow Mg_{III} + He + e$

The Mg$_{III}$ ions generated by double charge-exchange are removed from calculations. The collisional coupling for Mg$_I$ with other species is calculated according to elastic cross-section $10^{-15}$ cm$^2$. Mg$_{II}$ ion is taken to be strongly coupled due to Coulumb collisions. At the same time, the cotribution of Mg to overall heating/cooling is neglected because of its low abundance.

The relevance and limitations of the applied quasi-axisymmetric 2D approximation used here were studied in details in Khodachenko et al. (2015). In particular, within this approximation we disregard the Coriolis force and circularly (relative the planet-star line) average the centrifugal force. Such simplification is possible for sufficiently slowly rotating tidally locked planets. Note, that the planetary surface rotation velocity, which appears a symmetry breaking factor, is usually less than the considered typical velocities. Moreover, the Coriolis force doesn't change the energy of the moving material and its influence on the thermal mass loss rate is negligible. The centrifugal force is another factor, which also breaks the axial symmetry of the considered tidally locked planet-star system. It has maximum value in the ecliptic plane and turns to zero on the poles of rotation axis. The applied circular averaging around the planet-star line smooths this difference, which is also a relevant simplification for the considered processes.

Directly related with these geometry approximations is an expression for the gravitational stellar-planetary interaction which includes also the rotational effects. Averaged around the planet-star line it reads for a tidally locked system (Khodachenko et al. 2015) in the planet-based spherical coordinate system (R, θ, φ) with the star located at $\theta = 0$, at orbital distance $D$, as follows:

$$U = -\frac{GM_p}{R} \cdot \left(1 + \frac{1}{2}\frac{R^3}{R_{L1}^3}\frac{7\cos^2\theta - 1}{6}\right) \quad (A.5)$$

here $R_{L1} = D(M_p/3M_*)^{1/3}$ is the first Lagrange point.

One more limitation of the applied quasi-axisymmetric approximation for the tidally locked stellar-planetary system is the assumption of co-directionality of the relative velocity of SW flow and the ionizing and heating stellar radiation flux. In reality, the total relative velocity $\tilde{V}_{SW} = \sqrt{V_{SW}^2 + V_K^2}$ of an incoming SW, besides of the velocity $V_{SW}$ of the plasma wind itself, should include also the component of the planetary Keplerian orbital motion $V_K = \sqrt{GM_{st}/D}$, which for a Sun-like star is $\approx 30/\sqrt{D}$ km/s, where D is expressed in AU. Note that for extremely close exoplanets, the orbital component $V_K$ becomes even a dominating part of the total relative SW velocity vector $\tilde{V}_{SW}$, and in this case different directions of the SW and the stellar radiation flux break the symmetry of the problem.